\documentclass[aps,prl,10pt,twocolumn]{revtex4-2}
\usepackage{epsfig}
\usepackage{amsfonts}
\usepackage{mathtools}
\usepackage{amssymb}
\usepackage{enumitem}
\usepackage{amsmath}
\usepackage{graphicx}
\usepackage{amsmath}
\usepackage{amsthm}
\usepackage{extarrows}
\usepackage{bm}
\usepackage{hyperref}
\usepackage{mathrsfs}
\usepackage{bbm}
\usepackage{cancel}
\usepackage{hyperref}
\usepackage[dvipsnames]{xcolor}
\usepackage{accents}
\usepackage{relsize}
\usepackage[capitalize]{cleveref}
  
  \crefname{section}{Sec.}{Secs.}
  \crefname{appendix}{App.}{Apps.}
\usepackage{comment}

\usepackage{unicode-math}   
\usepackage{graphicx}
\usepackage{subcaption}

\usepackage{cleveref}

\allowdisplaybreaks

\usepackage{slashed}
\usepackage{xcolor}
\usepackage{hyperref}

\usepackage{amssymb,amsmath,graphics}   
\newcommand{\eq}{\begin{equation}}
\newcommand{\eqe}{\end{equation}}
\newcommand{\eqa}{\begin{eqnarray}}
\newcommand{\eqae}{\end{eqnarray}}

\newcommand{\td}{\text{d}}
\newcommand{\Stot}{S_{\G}}

\newcommand{\Tr}{\mathrm{Tr}}

\newcommand{\G}{\mathcal{G}}

\newcommand{\dsqcup}{\mathbin{\rotatebox[origin=c]{-90}{$\exists$}}}

\newcommand{\res}[1]{\mathop{\rm Res}\limits_{#1}}

\definecolor{myblue}{HTML}{00A1FF}

\makeatletter
\newcommand*{\balancecolsandclearpage}{%
  \close@column@grid
  \clearpage
  \twocolumngrid}
\makeatother

\begin{document}

\title{Cosmological Wavefunctions as Amplitudes: Dual Shuffle Factorization and Uniqueness from New Hidden Zeros}

\author{Yang Li$^1$}
\email{y.l.li@rug.nl}
\author{Laurentiu Rodina$^2$}
\email{laurentiu.rodina@gmail.com}

\affiliation{$\mbox{}^{1}$Van Swinderen Institute for Particle Physics and Gravity, University of Groningen,
 9747 AG Groningen, The Netherlands}

\affiliation{$\mbox{}^{2 }$Beijing Institute of Mathematical Sciences and Applications (BIMSA), Beijing, 101408, China}

\begin{abstract}

We show that cosmological wavefunctions in $\phi^n$ theories naturally generalize flat-space $\mathrm{Tr}(\phi^3)$ scattering amplitudes: via a simple map from tube variables to Mandelstam invariants, each wavefunction coefficient $\psi_{\mathcal{G}}$ becomes an on-shell amplitude-like object $\mathcal{A}_G$ associated with a generating graph $G$. At tree level these objects coincide with the Cachazo-He-Yuan construction based on Cayley functions that generalizes Parke-Taylor factors. We uncover new graph-based hidden zeros that extend and unify all known cosmological zeros. Based on this zero structure, we uncover a factorization principle dual to unitarity. Instead of factorization across poles, $A\to A_L\times A_R$, a zero at $p_{a\in G_L}\!\cdot\! p_{b\in G_R}=0$ factorizes the generating graph, $G\to G_L\times G_R$, and is equivalent to the shuffle decomposition $\mathcal{A}_G=\mathcal{A}_{G_L}\dsqcup \mathcal{A}_{G_R}$. Near-zero factorization is a simple consequence of this new structure. Using dual factorization, we show that locality together with the full set of hidden zeros uniquely fixes tree-level cosmological wavefunctions without assuming unitarity. We show that these zeros are equivalent to special enhanced large-$z$ behavior under Britto-Cachazo-Feng-Witten (BCFW) shifts, extending the zeros--BCFW correspondence beyond flat-space amplitudes. We also find evidence for further extensions of the zero structure and loop-level uniqueness. Our results show that cosmology provides a natural arena for on-shell methods and even reveals new structure in flat-space amplitudes.

\end{abstract}

\maketitle
\section{Introduction}
Scattering amplitudes and cosmological observables are traditionally derived from a Lagrangian by summing Feynman diagrams. In recent years, however, a complementary bootstrap viewpoint has emerged: instead of starting from microscopic dynamics, one attempts to reconstruct these quantities directly from general consistency principles.  This perspective has often revealed surprising simplicity and  structures that are completely hidden in the diagrammatic expansion (see \cite{Bern:2022jnl,Travaglini:2022uwo} for reviews).

A striking example of this philosophy is the discovery of \emph{hidden zeros} of scattering amplitudes: special kinematic loci where amplitudes vanish due to non-trivial cancellations among diagrams\cite{Arkani-Hamed:2023swr,DAdda:1971wcy,Arkani-Hamed:2023jry,Arkani-Hamed:2024fyd,Arkani-Hamed:2024nhp,Arkani-Hamed:2024vna,Arkani-Hamed:2024yvu,Bartsch:2024amu,Li:2024qfp,Dong:2024klq,De:2024wsy,Cheung:2024uhn,Guevara:2024nxd,Arkani-Hamed:2024tzl,Zhou:2024ddy,Du:2024soq,Li:2024bwq,Zhang:2024efe,Arkani-Hamed:2024nzc,Arkani-Hamed:2024pzc,GimenezUmbert:2025ech,Huang:2025blb,Early:2025ivr,Jones:2025rbv,Chang:2025cqe,Paranjape:2025wjk,Feng:2025ofq,Cao:2025lzv,Bartsch:2025loa,Brauner:2025rzv,Backus:2025njt,Figueiredo:2025fnr,Berman:2025owb,Cheung:2025tbr,Li:2025suo,Bartsch:2025mvy,Feng:2025dci,Zhang:2025zjx,Zhou:2025tvq,Backus:2025orx,Dong:2025lrf,Carrolo:2025jca,Azevedo:2025vxo,Zhang:2026dcm,CarrilloGonzalez:2026lnu,Bartsch:2026mnl,Ruan:2026xyd,Basile:2026gnd}. Unlike singularities and residues, whose origin is understood through locality and unitarity, these zeros have no clear physical interpretation. They are invisible term by term yet extremely constraining: in several flat-space theories they uniquely determine amplitudes and even loop integrands \cite{Rodina:2024yfc,Backus:2025hpn}. Surprisngly, locality and unitarity need not be assumed but emerge from the zero structure itself. This suggests the zeros reflect a deeper and still mysterious organizing principle of quantum field theory. 

In parallel, cosmological observables have begun to be studied from a similar bootstrap viewpoint \cite{Arkani-Hamed:2018kmz,Baumann:2019oyu,Baumann:2020dch,Baumann:2022jpr}, and dressing rules that relate in-in correlators with flat space scattering amplitudes have been discovered \cite{Chowdhury:2025ohm,Chowdhury:2026upp}. The late-time cosmological wavefunction is the natural analog of the $S$-matrix in cosmology: its coefficients encode the probabilities of field configurations in the early universe and determine late-time correlators. For scalar theories, these coefficients admit a simple combinatorial representation as sums over maximal tubings of graphs, with each connected subgraph, or tube, carrying a variable $S_I$ \cite{Arkani-Hamed:2017fdk,Arkani-Hamed:2018bjr,Benincasa:2019vqr} (see also \cite{Benincasa:2022gtd} for a review). Unexpectedly, they exhibit many of the same structural features as scattering amplitudes, including rich singularity structure, connections to positive geometries such as cosmological polytopes and the cosmohedron \cite{Arkani-Hamed:2024jbp} (see also \cite{Glew:2025otn,Glew:2025ugf,Figueiredo:2025daa,Glew:2025mry,Arkani-Hamed:2025mce,Ardila-Mantilla:2026cbo,Glew:2026von,McLeod:2026jpz,Fu:2026dqb}), and even cosmological versions of hidden zeros \cite{De:2025bmf}. At the same time, flat-space amplitudes arise from distinguished boundaries, or ``scattering facets,'' of these geometries, suggesting a deep and bidirectional link between amplitudes and cosmological wavefunctions.

In this Letter we will show this connection is much stronger than previously appreciated.

A natural bootstrap question is whether cosmological zeros are powerful enough to uniquely determine cosmological wavefunctions, much as hidden zeros do for flat-space amplitudes. This question is important for two reasons. Practically, uniqueness would provide a direct route to constructing cosmological observables. Conceptually, it identifies which principles are fundamental, and which structures --- such as locality or unitarity --- may instead emerge from them, as previously observed in various flat space examples \cite{Arkani-Hamed:2016rak,Rodina:2018pcb,Rodina:2016jyz,Rodina:2016mbk,Carrasco:2019qwr,Rodina:2020jlw}. We find, however, that the known cosmological zeros are not sufficient to fix wavefunctions for general graphs. But this failure is informative: if we expect uniqueness to hold, it signals that cosmological wavefunctions must obey additional hidden structure beyond what was previously known. 

Motivated by the hidden zero-BCFW scaling equivalence \cite{Rodina:2024yfc}, to uncover this structure using on-shell BCFW shifts \cite{Britto:2004ap,Britto:2005fq,Carrasco:2019qwr} as probes, we introduce a simple kinematic map from tube variables to Mandelstam invariants
\eq
S_I \;\to\; s_I=\Bigl(\sum_{i\in I}p_i\Bigr)^2,
\label{map}
\eqe
where $I$ denotes a subset of vertices of $\mathcal{G}$. The map \eqref{map} turns each $n$-point wavefunction coefficient $\psi_{\mathcal{G}}$ into an \emph{on-shell} $(n+1)$-point amplitude-like object $\mathcal{A}_G$ associated with its generating graph $G\equiv \mathcal{G}\cup {v_{n+1}}$. As will be explained, the map implicitly adds an auxiliary vertex to the graph $\mathcal{G}$. Remarkably, at tree level, these objects land precisely on the Cachazo-He-Yuan (CHY) Cayley construction \cite{Gao:2017dek}. 

In this language, new cancellations and enhanced large-\(z\) behavior appear, signaling the existence of new zeros. We thus uncover a graph-based extension of hidden zeros that extends and unifies the known cosmological zeros. Concretely, whenever two vertices \(i,j\) split \(G\) into subgraphs \(G_L,G_R\), we find a wavefunction zero when setting
\eq
p_{a\in G_L}\!\cdot p_{b\in G_R}=0.
\eqe

The search for uniqueness from zeros then leads to a more surprising discovery. By analyzing how the new zeros constrain a local ansatz, we find that they imply a factorization principle dual to unitarity: zeros factorize the generating graph into subgraphs, and on the kinematic side this is realized as a shuffle product of lower-point objects
\eq
\textrm{zero on } p_{a\in G_L}{\cdot} p_{b\in G_R}{=}0 \ \Leftrightarrow \ \mathcal{A}_G = \mathcal{A}_{G_L}{\dsqcup} \ \mathcal{A}_{G_R} \,.
\eqe
By contrast, ordinary unitarity factorizes amplitudes across poles into a left--right product of subamplitudes
\eq
\textrm{unitarity on } P^2=0:\qquad \mathcal{A} = \mathcal{A}_L \times \mathcal{A}_R \, .
\eqe
Furthermore, we find this shuffle structure directly implies the near-zero factorization \cite{Arkani-Hamed:2023swr}. Thus the quest for cosmological uniqueness does more than provide a bootstrap construction: it uncovers a new physical meaning of zeros and a new organizing principle shared by amplitudes and cosmological wavefunctions.

\medskip

To summarize, our main results are as folllows:

\textbf{Kinematic map and CHY realization.} We introduce a simple kinematic map from tube variables to Mandelstam invariants that sends each cosmological wavefunction coefficient to an amplitude-like object. At tree level, these objects coincide precisely with the CHY construction based on Cayley functions.

\textbf{New graph-based hidden zeros.} In these variables, we uncover a graph-based generalization of hidden zeros that both extends and unifies the previously known cosmological zeros.

\textbf{Zeros as dual factorization.} We show that these zeros admit a factorization principle dual to unitarity: an object satisfies the zero conditions precisely when it can be written as a shuffle product of lower-point objects associated with the factor graphs.

\textbf{Uniqueness from dual factorization.} This dual factorization is strong enough to uniquely fix all tree-level cosmological wavefunctions from locality and zeros alone.

\textbf{Equivalence to BCFW scaling.} We show that these graph-based zeros are exactly equivalent to the corresponding enhanced large-$z$ behavior under BCFW shifts.

We conclude by briefly highlighting observations for future work. This includes hints of an even larger class of zeros, and a potential extension of uniqueness to arbitrary loop level.

\section{Review: The cosmological wavefunction}

We study the late-time wavefunction of a conformally-coupled scalar with polynomial interactions in an FRW background. Here we introduce the cosmological wavefunctions, which are the main objects of study in this work. A more detailed review is given in Appendix 1.

After the standard conformal map, and restricting to the flat-space case with constant couplings, the relevant boundary observable is the vacuum wavefunction
\begin{align}
    \Psi[\Phi] &= \int_{\phi(-\infty(1-i\epsilon))}^{\phi(0)=\Phi} \mathcal{D}\phi\, e^{iS[\phi]} \, ,
\end{align}
whose perturbative expansion defines the wavefunction coefficients $\psi_n$. These coefficients are the basic observables of interest in this work: they determine late-time correlators and provide the cosmological analogue of scattering data. We focus on the flat-space case with constant couplings, since more general cosmological observables can be obtained  by a simple integral transform \cite{Arkani-Hamed:2023kig,Arkani-Hamed:2023bsv,De:2023xue,De:2024zic}.

\begin{figure}[h] 
   \centering
   \includegraphics[width=\linewidth]{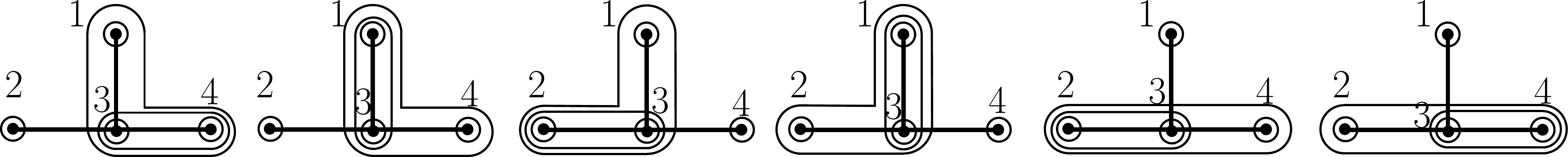} 
   \caption{All maximal tubings of a 222-star.}
   \label{fig: tubing examples}
\end{figure}

A remarkable feature of these coefficients is that they admit a purely combinatorial description \cite{Arkani-Hamed:2017fdk}. For a graph $\G$, the corresponding coefficient can be written as a sum over maximal tubings $T$ of $\G$,
\begin{align}
    \psi_{\G}=\sum_{T}\prod_{\tau\in T}\frac{1}{S_{\tau}}
    =\frac{1}{\Stot\mathop{\prod}\limits_{\tau\in \{\text{1-tubes}\}}S_{\tau}}\,
    \widetilde{\psi}_{\G} \, ,
    \label{eq: wavefunction coefficients}
\end{align}
where $\widetilde{\psi}_{\G}$ denotes the stripped coefficient after removing the universal total-energy and 1-tube factors. This stripped object is what we will refer to as the wavefunction coefficient. For simplicity of notation, we will refer to the stripped wavefunction also as $\psi_{\mathcal{G}}$. For example, for the $222$-star in FIG.~\ref{fig: tubing examples} one finds $\psi_{\mathcal{G}}(1234)$ is given by
\eq\label{222}
\frac{1}{S_{34}S_{134}}+\frac{1}{S_{13}S_{134}}+\frac{1}{S_{23}S_{132}}+\frac{1}{S_{13}S_{132}}+\frac{1}{S_{23}S_{234}}+\frac{1}{S_{34}S_{234}} \, .
\eqe
These stripped coefficients were recently shown to vanish on several classes of loci, including parametric, wavefunction, and factorization zeros \cite{De:2025bmf}. Our aim in this work is to connect ${\psi}_{\G}$ to amplitudes and explore their combinatorial structure from an on-shell perspective.

\section{Kinematic Wavefunctions}
We begin by analyzing the wavefunction coefficients through the kinematic map \eqref{map}.
Under this map, together with on-shell conditions for the momenta $p_i^2=0$ and momentum conservation $\sum_{i=1}^{n+1}p_i=0$, the wavefunction ${\psi}_{\G}$ associated to a graph $\G$ with $n$ vertices becomes $(n+1)$-point amplitude-like objects associated with graphs $G$, which we refer to as $\mathcal{A}_{G}$, \emph{kinematic wavefunctions}. Here the graph $G$ is the original wavefunction graph,plus an auxiliary vertex $n+1$, connected to all valency 1 vertices in $\G$. 

These on-shell objects $\mathcal{A}_G$ are simply sums over particular permutations of Tr$(\phi^3)$ ordered amplitudes, together with a restricted set of propagators, determined by the underlying graph. Ordinary ordered amplitudes correspond to chain graphs (which therefore have an inherent cyclic symmetry), while more general graphs give rise to more general kinematic wavefunctions. 
\begin{figure}[h] 
   \centering
   \includegraphics[width=\linewidth]{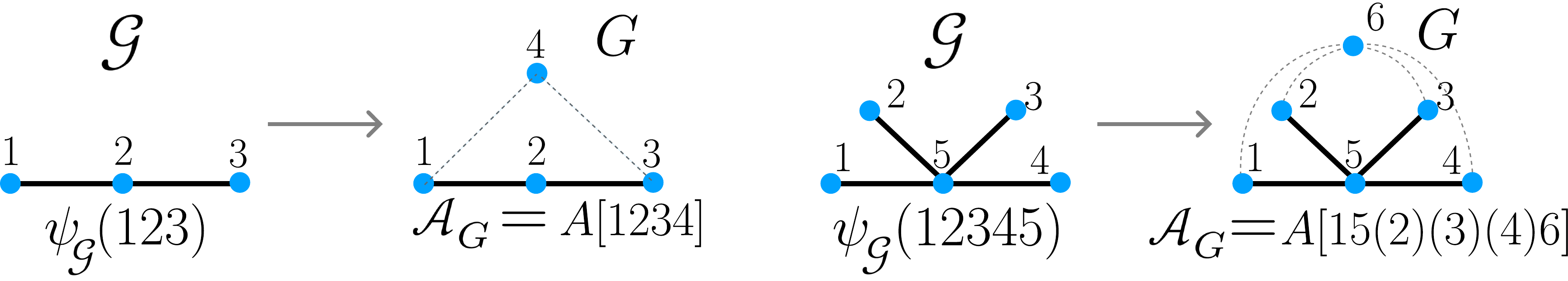} 
   \caption{The 3-chain $\psi(123)$ is mapped to an ordered Tr($\phi^3$) 4-point amplitude $A[1234]$. Fixing vertex 1, the 2222-star is mapped to $A[15(2)(3)(4)6]$, which involves 6 different orderings, the permutations of $\{2,3,4\}$.}
   \label{fig:simple}
\end{figure}
As a first example, consider the wavefunction for the 3-chain 
\eq
\psi(123)=\frac{1}{S_{12}}+\frac{1}{S_{23}} \, .
\eqe
Under the map \eqref{map}, these two terms map precisely to the two Feynman diagrams of the ordered amplitude $A[1234]$, with momentum $p_4$ introduced implicitly via momentum conservation. Graphically, the vertex $4$  is understood to connect to the two endpoints, $1$ and $3$, as shown in FIG.~\ref{fig:simple}. The position of the $(n\!+\!1)$-th vertex can be inferred from BCFW scaling of the on-shell object $\mathcal{A}_G$, but its origin also follows naturally from the CHY construction: it is an extra vertex required by ${\rm SL}(2,\mathbb C)$ invariance.

As a more instructive example, consider the $222$-star $\psi_{\mathcal{G}}(1234)$ of Eq.~\eqref{222}. Under the map \eqref{map}, all six terms are reproduced by the sum of two ordered five-point amplitudes,
\eq\label{G_to_A}
\mathcal{A}_G[13(2)(4)5]= A[132|45]+A[134|25]\, .
\eqe
Here $2|4$ denotes the exclusion of diagrams containing the pole $s_{24}$, in accord with the star graph, where vertices $2$ and $4$ are not adjacent and hence no tube $S_{24}$ exists. As before, the $(n+1)$-th vertex is implicit and is taken to be adjacent to all valency-one vertices.

In fact, any graph maps to a sum of color-ordered Tr$(\phi^3)$ amplitudes, even if the graph itself contains higher-valency vertices. The cubic structure is dictated not by the valency of the original graph, but by the structure of maximal tubings. The graph topology only constrains the external partial orderings of the resulting Tr$(\phi^3)$ amplitudes. For instance, the 2222-star in FIG.~\ref{fig:simple}  maps to a sum over six permutations of ordered Tr$(\phi^3)$ amplitudes. For an even more general example, the kinematic wavefunction in FIG.~\ref{fig:greenblob} can be written as a sum of four orderings
\begin{align}
\nonumber &\mathcal{A}_G[13(2)(45(6)(78))9]=
A[132|456|789]+\\&+A[132|4578|69]+A[13456|78|29]+A[134578|6|29]
\end{align}

In this way, the kinematic map places cosmological wavefunctions and scattering amplitudes in a common on-shell language. The objects $\mathcal{A}_G$ are generalized ordered amplitudes associated with arbitrary graphs, reducing to ordinary color-ordered amplitudes for chain-graphs. As we will see, this viewpoint reveals genuinely new structure that is obscured in either description separately.

Importantly, this allows us to focus mostly on ordered amplitudes - cosmological wavefunctions simply involve summing over different orderings, ensuring that many amplitude level results carry over to wavefunctions.

\subsection{Kinematic wavefunctions from CHY}

Unexpectedly, we find that the kinematic map from cosmological wavefunctions coincides precisely with the CHY Cayley construction associated with labeled tree graphs \cite{Gao:2017dek}.

For ordinary color-ordered amplitudes, the CHY half-integrand is the Parke--Taylor factor, corresponding to a chain graph \cite{Cachazo:2013gna,Cachazo:2013hca,Cachazo:2013iea,Cachazo:2014nsa,Cachazo:2014xea}. In \cite{Gao:2017dek}, this was generalized to an arbitrary tree graph $\G$ through the \emph{Cayley function}, obtained by assigning a factor $\frac{1}{\sigma_{vv'}}$ to each edge $e(v\leftrightarrow v')\in \G$. For example, the $222$-star in FIG.~\ref{fig: tubing examples} is associated with
\begin{align}
    \frac{1}{\sigma_{1,3}\sigma_{2,3}\sigma_{3,4}} \, .
    \label{eq: Cayley function 222}
\end{align}
The CHY integral with integrand given by the square of the Cayley function then produces precisely the corresponding kinematic wavefunction, and by extension, the cosmological wavefunction itself.

The expression in \eqref{eq: Cayley function 222} is the gauge-fixed form of the half-integrand. To define the CHY integrand covariantly, one must restore the correct ${\rm SL}(2,\mathbb C)$ weight~2 at every puncture. 
This requires introducing one additional marked point $\sigma_{n+1}$ and multiplying by a compensating factor.

For the $222$-star, the fully ${\rm SL}(2,\mathbb C)$-covariant Cayley function becomes
\begin{align}
    \frac{\sigma_{3,5}}{\sigma_{1,3}\sigma_{2,3}\sigma_{3,4}\sigma_{1,5}\sigma_{2,5}\sigma_{4,5}} \, .
    \label{eq: Cayley function 222 covariant}
\end{align}
The extra factors involving $\sigma_{5}$ show that the auxiliary vertex $v_{n+1}$ is naturally treated as adjacent to all univalent vertices.

\section{Wavefunction blob zeros }
Next, we will show that such wavefunctions satisfy the natural generalization from amplitude ``hidden zeros" to wavefunction ``blob zeros", which are associated to more general graphs.

For amplitudes, a zero can be found by specifying two non-adjacent legs $i$ and $j$ (see also \cite{Cao:2024gln,Cao:2024qpp} from the universal splitting perspective). Given a fixed cyclic ordering (equivalently, a chain graph), this defines two sets $A$ and $B$ as $(A,i,B,j)$, such that any leg in $A$ is not adjacent to any leg in $B$. Then 
\eq\label{zero_def}
p_a\cdot p_b=0,\textrm{ for all } a\in A, b\in B
\eqe
is a zero locus where the amplitude non-trivially vanishes. As shown in Ref.~\cite{Rodina:2024yfc}, the same pair of vertices $i,j$ also determines the BCFW shift for which the zero condition is exactly equivalent to enhanced large-$z$ scaling.

We find that a cosmological zero is then the natural extension of this construction to a general graph. We simply need to find subsets $A$, $B$ that are mutually not-adjacent.

\begin{figure}[h] 
   \centering
   \includegraphics[width=0.9\linewidth]{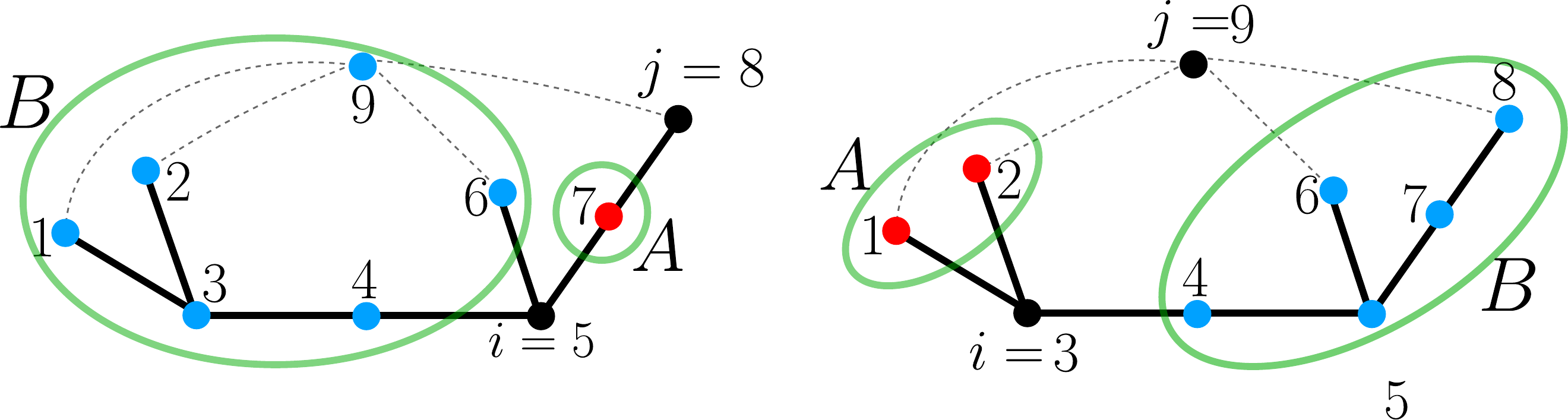} 
   \caption{Left: A blob zero uniquely defined by $\partial A=\partial B=\{5,8\}$. Right: The choice $i,j=\{3,9\}$ splits the graph into the three branches of vertex 3; $A=\{1,2\}, B=\{4,5,6,7,8\}$ gives one of the zeros. The colored notation corresponds to (${\color{red} A},i,{\color{myblue} B},j$).}
   \label{fig:greenblob}
\end{figure}

First, let us define the external boundary of a subgraph $A\subset {G}$
\eq
\partial A := \{\, v\in G\setminus A : v \text{ is adjacent to some } a\in A \,\}
\eqe
Then, for any $A$ whose external boundary contains exactly two elements, $\partial A=\{i,j\}$, define $B$ as the non-adjacent complementary set $B=G\setminus (A\cup \partial A)$. Note that $\partial A=\partial B=\{i,j\}$. Then the conditions in Eq.~\eqref{zero_def} are a zero locus for the wavefunction $\mathcal{A}_G$, which we call a \emph{blob zero}.
An example is shown in the LHS of FIG.~\ref{fig:greenblob}.

Equivalently, a valid blob zero is defined whenever the vertices $i,j$ cut the graph into at least two subgraphs. In general, $i,j$ may cut the graphs into $k$ subgraphs $V_i$, with $\partial V_i=\{i,j\}$. It is easy to see this can only happen if $\textrm{valency}[i]=k$ and $j={n+1}$. In this case, a zero exists even if $A$ or $B$ are not connected subgraphs, but any disjoint union of the $V_i$. An example of this more general splitting is shown in the RHS of FIG.~\ref{fig:greenblob}. This cutting perspective will become useful when we next discuss dual factorization.

Thus blob zeros provide the natural extension of hidden zeros from ordinary amplitudes to cosmological wavefunctions. In Appendix 2 we show the blob zeros unify and extend the previously known cosmological zeros.

\section{Dual factorization}
Due to unitarity and the optical theorem, poles are governed by factorization: when an intermediate channel goes on shell, the amplitude splits into lower-point amplitudes. What is the analogous principle behind zeros?

In this section we prove that a wavefunction $\mathcal{A}_G$ satisfies a blob zero $A,B$ if and only if it can be decomposed into a shuffle product schematically of the form
\eq
\mathcal{A}_G\rightarrow \mathcal{A}_{G_A}\dsqcup \mathcal{A}_{G_B}
\eqe

A first clue comes from the search for uniqueness. Empirically, when one imposes a zero on increasingly complicated local wavefunction ans\"atze associated with general graphs, the number of unfixed coefficients is found to factorize as
\eq
\#(G)_{(A,B)\text{-zero}}=\#(G_A)\times \#(G_B)\, .
\label{factorN}
\eqe
where $G_A$ and $G_B$ are subgraphs determined by the zero. Their precise meaning will be given later.

Therefore, we find the zero does not directly factorize the amplitude in the usual sense; rather, it appears to factorize the underlying generating graph. This is why we refer to the resulting structure as \emph{dual factorization}.

For ordinary amplitudes, this structure is already implicit in the $D$-subset construction of Ref.~\cite{Rodina:2024yfc} and resonant with some results in \cite{Li:2024bwq}. Consider a local color-ordered ansatz $\mathcal B_n$, and a zero specified by two sets $A$ and $B$ with common boundary $\partial A=\partial B=\{i,j\}$. 
Ref.~\cite{Rodina:2024yfc} showed that $\mathcal B_n$ satisfies the zero if and only if it can be decomposed into $D$-subsets such that each $D$-subset vanishes independently on the zero locus, and all terms within a given $D$-subset carry the same coefficient.

Let us consider a 5 point example, and the local ansatz
\eq\label{5pt}
\mathcal{B}_5=\frac{c_1}{s_{12}s_{123}}+\frac{c_2}{s_{23}s_{234}}+\frac{c_3}{s_{34}s_{12}}+\frac{c_4}{s_{123}s_{23}}+\frac{c_5}{s_{234}s_{34}}
\eqe
After imposing the zero condition $A=\{1\}$, $B=\{3,4\}$, we find the constraints $c_1=c_2=c_4$ and $c_3=c_5$. The two left-over coefficients correspond to two $D$-subsets
\eqa\label{5p_Dsubset}
D_1&=&c_1\left(\frac{1}{s_{12}s_{123}}+\frac{1}{s_{23}s_{123}}+\frac{1}{s_{23}s_{234}} \right)\\
D_2&=&c_3\left(\frac{1}{s_{12}s_{34}}+\frac{1}{s_{34}s_{234}}\right)
\eqae
each of which independently vanishes under the zero, and with all terms having equal coefficient, as stated above.

\begin{figure}[h!]
    \centering
    \includegraphics[width=0.6\linewidth]{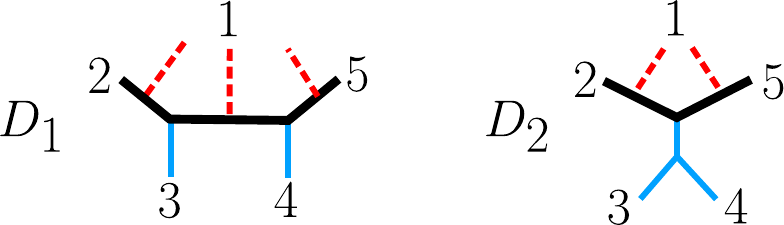}
    \caption{5-point amplitude $D$-subsets for zero $p_1\cdot p_3=p_1\cdot p_4=0$. Dashed lines represent different choices of inserting leg 1. The colored notation corresponds to (${\color{red} A},i,{\color{myblue} B},j$). Each D-subset simply corresponds to the shuffle between two diagrams, the top one corresponding to $(i,A,j)$, the bottom one to $(i,B,j)$.}
    \label{fig: 5pt D-subsets}
\end{figure}

Importantly, the two subsets have a simple geometric meaning. They can be generated by adding leg 1 to the two different 4 point diagrams, see FIG.~\ref{fig: 5pt D-subsets}.

The construction for a general zero is completely analogous. One removes one side of the zero, say the block $A$, leaving the lower-point ordered amplitude $(i,B,j)$, and then reinstates the removed block in all order-preserving ways. When $A$ contains more than one leg, one must also sum over all allowed internal topologies of that block.

This is precisely a \emph{shuffle product}. The counting formula
\eqref{factorN}, together with the symmetry between the two sides of the zero,
indicates that the natural lower-point building blocks are the diagrams with
external orderings \((i,A,j)\) and \((i,B,j)\). A \(D\)-subset is then obtained
by shuffling these two ordered pieces along the common line \(i\!-\!j\), while
preserving the internal ordering within each factor:
\eq
D=\Gamma_{i,A,j}\dsqcup \Gamma_{i,B,j}\, .
\eqe
Here \(\dsqcup\) denotes the sum over all such order-preserving interleavings between diagrams $\Gamma$.

A general local ansatz is obtained by summing over all diagrams
\(\Gamma_{i,A,j}\) and \(\Gamma_{i,B,j}\) with independent coefficients.
Summing over all \(D\)-subsets then gives the condition for a general local
ansatz to satisfy the zero,
\eq
 \mathcal B_n \text{ satisfies $(A,B)$-zero }
 \Longleftrightarrow
\mathcal B_n = \mathcal B_{(i,A,j)} \dsqcup \mathcal B_{(i,B,j)} \, .
\label{eq:amp-shuffle}
\eqe
The use of \(\mathcal B\) is intentional: this statement is more general than
factorization into lower-point amplitudes. The only requirement is a shuffle
decomposition into lower-point local building blocks, which need not themselves
be amplitudes or wavefunctions. The stronger case \(\mathcal B=\mathcal A\)
will be discussed in the next section.

\begin{figure}[h!]
    \centering
    \includegraphics[width=1\linewidth]{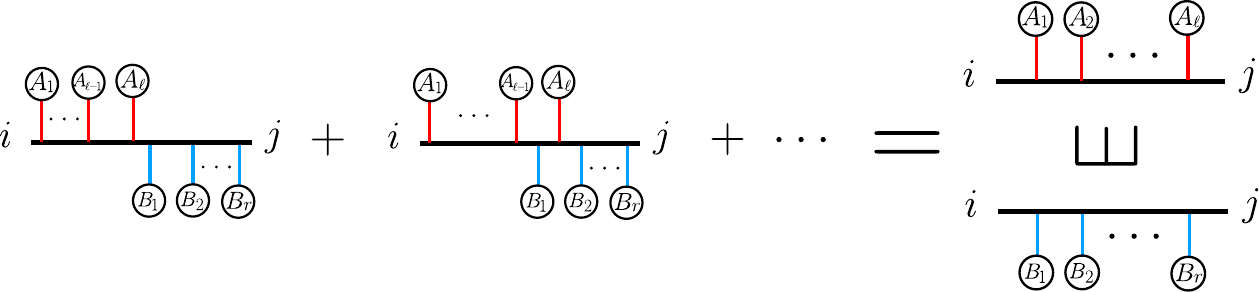}
    \caption{LHS: A $D$-subset from $\mathcal{B}_n$. RHS: The shuffle product of two lower-point diagrams, one from $\mathcal{B}_{(i,A,j)}$ and the other from $\mathcal{B}_{(i,B,j)}$, generating the subset on the LHS. The colored notation corresponds to (${\color{red} A},i,{\color{myblue} B},j$). $A_1\cup\dots \cup A_{\ell}=A$, and  $B_1\cup \dots \cup B_r=B$.}
    \label{fig: shuffle product of amplitude}
\end{figure}

Thus, for amplitudes (or local objects), a zero is realized not by an ordinary product of lower-point amplitudes (or local objects), but by an exact shuffle product of them. A shuffle product of a diagram from $\mathcal{B}_{(i,A,j)}$ and the other from $\mathcal{B}_{(i,B,j)}$, is illustrated in FIG.~\ref{fig: shuffle product of amplitude}. Eq.\eqref{eq:amp-shuffle} includes all combinations of shuffle product from $\mathcal{B}_{(i,A,j)}$ and $\mathcal{B}_{(i,B,j)}$.

\subsection{Dual factorization for wavefunctions}
\begin{figure}
    \centering
    \includegraphics[width=0.8\linewidth]{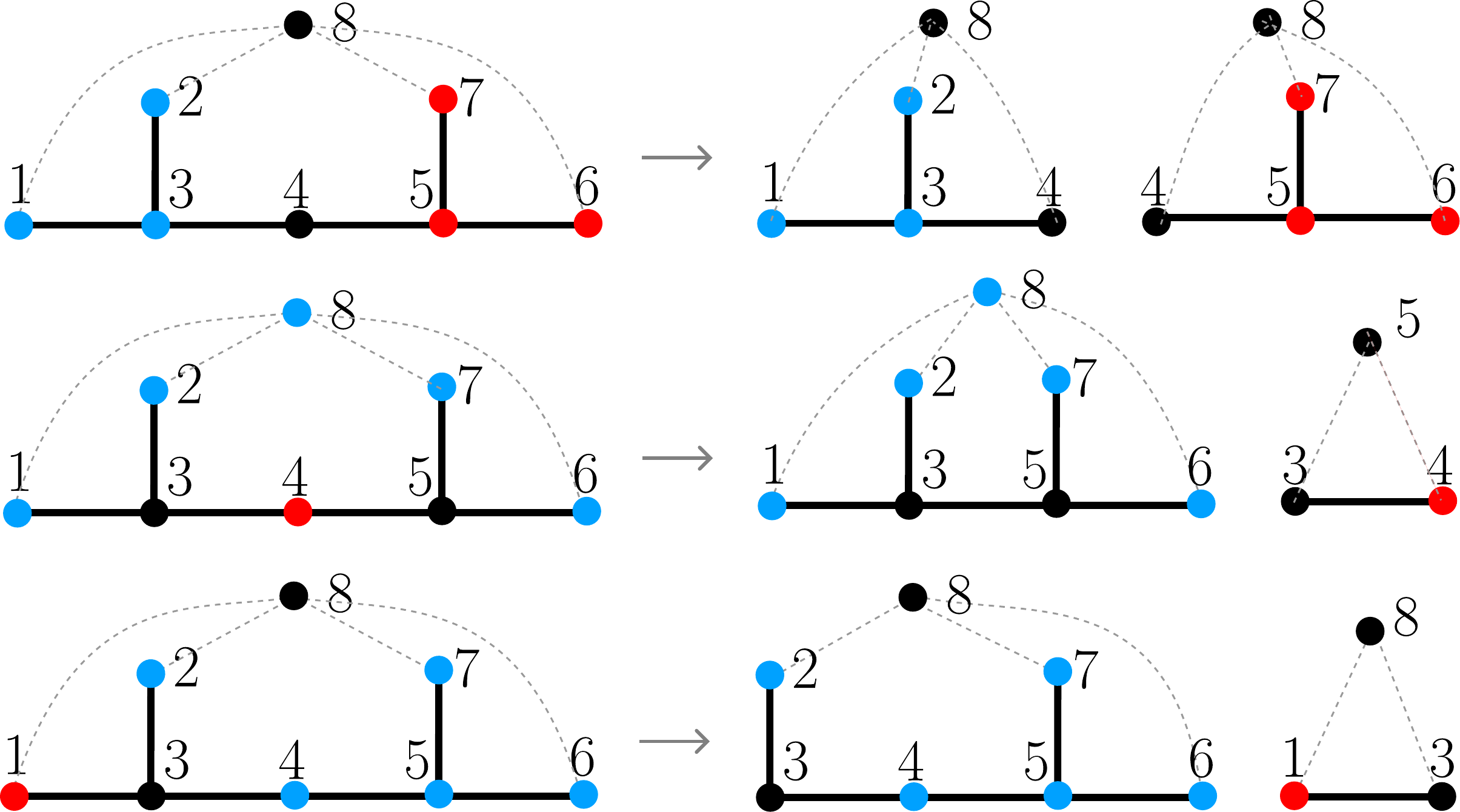}
    \caption{Examples to illustrate the rules in \eqref{GA}. The colored notation corresponds to (${\color{red} A},i,{\color{myblue} B},j$). }
    \label{fig: GA GB examples}
\end{figure}

For cosmological wavefunctions, the same shuffle structure admits a natural graph interpretation. A zero on $(A,B)$ now factorizes the generating graph $G$ itself into two factor graphs $G_A$ and $G_B$. If
$\partial A=\partial B=\{i,j\}$, the factor graphs are then obtained by the following rule:
\begin{align}
\label{GA}
G_A = 
\begin{cases}
A\cup\{i,j\} & \exists \, S^1\,|\, i,j \in S^1\subset A\cup\{i,j\}\, , \\
A\cup\{i,j\}\cup e(i\leftrightarrow j)  &  \text{otherwise} \, .
\end{cases}
\end{align}
A new edge $e=(i\leftrightarrow j)$ between $i,j$ is added if there is not a circle subgraph $S^1$ that consisting of both $i,j$ as well as some vertices in $A$; similarly for $G_B$. Nontrivial examples can be found in FIG.~\ref{fig: GA GB examples}.

The corresponding local ansatz for a wavefunction must then take the form
\begin{equation}
\mathcal B_G = \mathcal B_{G_A} \dsqcup \mathcal B_{G_B}\, .
\label{eq:wvf-shuffle}
\end{equation}
 For amplitudes this reduces to the simple statement that a chain is cut into two smaller chains and shuffled back together; for cosmological wavefunctions the process is identical, only involving more general graphs. A complete proof for dual factorization for wavefunctions is given in Appendix 3.

\begin{figure}[h!] 
   \centering
   \includegraphics[width=\linewidth]{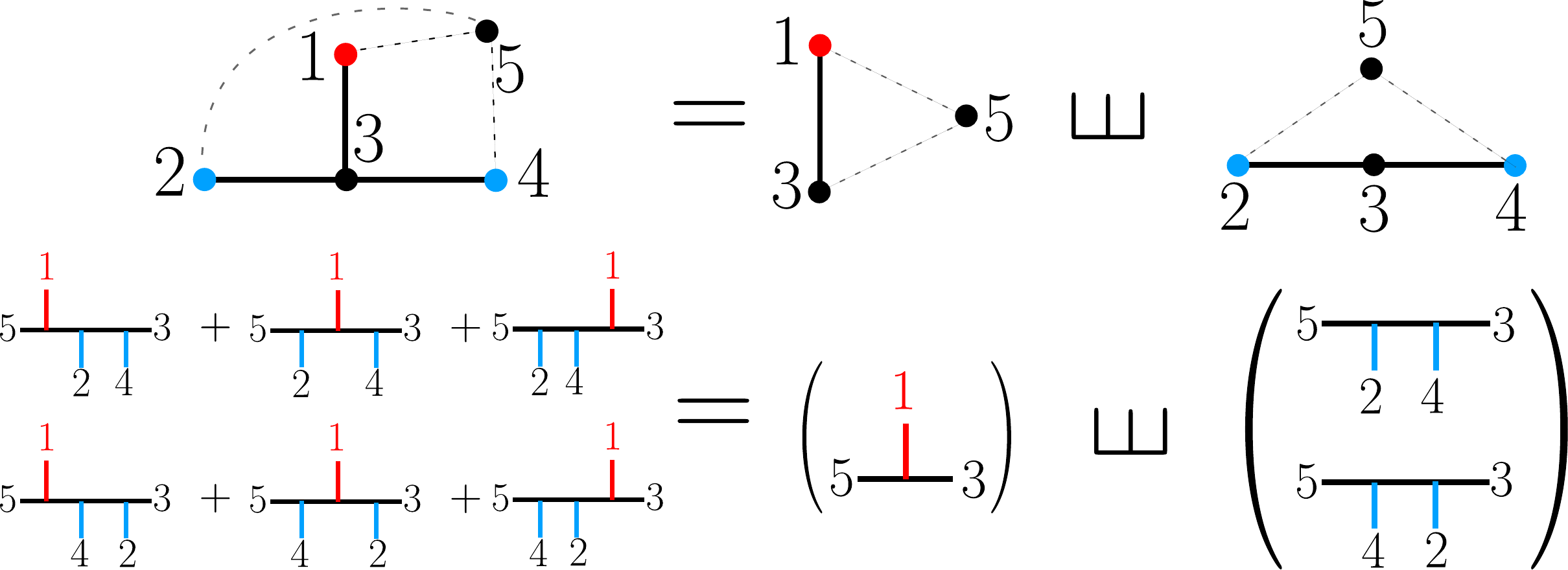} 
   \caption{Graph dual factorization. Map from graph product is shuffle product of diagrams. Amplitude satisfies ({\color{red} 1}), ({\color{myblue}24}) zero iff it can be written as shuffle product of a 3-point amplitude and a two 4-point ones. The colored notation corresponds to (${\color{red} A},i,{\color{myblue} B},j$).}
   \label{fig:example}
\end{figure}

Hidden zeros therefore encode a structure strikingly parallel to ordinary factorization. Geometrically, the two are dual. Ordinary factorization cuts diagrams \emph{across} a propagator and isolates the residue on a physical pole, whereas a zero cuts diagrams \emph{along} a line connecting the boundary pair \(i,j\), splitting the external data into two ordered pieces whose shuffle reconstructs the full object.

\subsection{Near-zero factorization from dual factorization}
An important refinement of the dual-factorization picture is that the $D$-subsets themselves also obey near-zero factorization. Here we briefly motivate the result, and leave more details to Appendix 4. Each $D$-subset is a shuffle product of two lower-point diagrams, and it can be shown that relaxing one zero condition  collapses the shuffle into an ordinary product. Concretely, on a near zero defined by $A,B$ and $c_{ij}\neq 0$, each $D$-subset constructed by the shuffle of two diagrams $\Gamma_A$ and $\Gamma_B$ 
\eq
D=\Gamma_A \dsqcup \Gamma_B
\eqe
factorizes as
\begin{equation}
D \;\longrightarrow\; \frac{c_{ij}}{s'_{A,j}s'_{j,B}}\times \widetilde{\Gamma}_A \times\widetilde{\Gamma}_B  \, ,
\end{equation}
and summing over all $D$-subsets we obtain
\begin{equation}
A(i,A,j,B)\;\longrightarrow\;\frac{c_{ij}}{s'_{A,j}s'_{j,B}}\times \widetilde{A}(i,A,j)\times \widetilde{A}(i,B,j)\, .
\end{equation}
here by $s'$ and $\widetilde{\Gamma}$ we simply mean the kinematics are evaluated on the near-zero conditions. This is the simple origin of the deformed kinematics that appears in near-zero factorization. 
Thus near-zero factorization is already encoded in dual factorization, and the same conclusion applies to cosmological wavefunctions.

\section{Uniqueness from hidden zeros} 

With dual factorization in hand, the proof of uniqueness becomes straightforward, which was the original motivation for this work. Since imposing a zero on $(A,B)$ requires the ansatz $\mathcal{B}_G$ to decompose according to Eq.~\eqref{eq:wvf-shuffle}, this proves the number of coefficients left unfixed is given by Eq.~\eqref{factorN}.
Iterating such factorizations eventually reduces any graph to a product of $2$-chain seeds, each of which carries only a single overall free coefficient. It follows that a sufficient set of zeros uniquely determines any wavefunction, up to overall normalization. The minimal number of zeros that guarantee uniqueness is simply obtained as the minimal number of operations needed to fully factor a graph into 2-chains. In general, this is $|\textrm{vertices}(G)|-3$, matching the known result for amplitude uniqueness. 

For example, for 222-star, there are three blob zeros: $(A,B)=(\{1\},\{2,4\})$,  $(A,B)=(\{2\},\{1,4\})$ and  $(A,B)=(\{4\},\{1,2\})$. Imposing the first zero leads to a dual factorization shown in FIG.~\ref{fig:example}, and the second zero leads to a further factorization of the 3-chain labeled by 2,3,4 into two 2-chains, labeled by $2,3$ and $3,4$. Since the graph is already dual-factorized into 2-chain seeds, uniqueness is already reached before imposing the third zero. Significantly more complicated examples are given in Appendix 3, FIGS. \ref{fig: 5-chain graph reduction},\ref{fig: 234 graph reduction},\ref{fig: 222-2222-222 graph reduction}. This includes an 11-vertex 222-2222-222 triple-star graph, which has 82320 initial unfixed coefficients, but is uniquely fixed by 8 zero conditions.

\subsection{Unitarity from dual factorization}
However, dual factorization yields more than uniqueness up to normalization. We showed above that any local object admitting a shuffle decomposition into two lower point objects automatically satisfies the corresponding zero. One may therefore elevate this from a consequence of zeros to a fundamental principle, in direct analogy with unitarity: instead of merely imposing the zero, or the generic shuffle condition between arbitrary local objects of Eq.~\eqref{eq:amp-shuffle}, one demands factorization into the actual lower-point \emph{wavefunctions},
\eq\label{dual_unitarity}
\mathcal{A}_G=\mathcal{A}_{G_L}\dsqcup \mathcal{A}_{G_R}\, .
\eqe
This has an important advantage. Purely kinematic uniqueness theorems  leave an overall normalization unfixed. By contrast, once dual factorization is imposed as the nonlinear principle in Eq.~\eqref{dual_unitarity}, the normalization of $\mathcal{A}_G$ is fixed recursively by the couplings of the lower-point factors. In this sense, dual factorization provides precisely the dynamical input that ordinary uniqueness arguments lack.

Furthermore, unlike ordinary factorization, which fixes one pole at a time, dual factorization determines the full object through an exact shuffle recursion.  Since unitarity is ultimately the expression of probability conservation, this suggests that hidden zeros may likewise be governed by a comparably fundamental physical principle.

\section{Equivalence to BCFW scaling}

Finally, we show the hidden-zero structure of cosmological wavefunctions is closely tied to their behavior under BCFW shifts, in direct parallel with scattering amplitudes.

For ordinary amplitudes, BCFW deformations often exhibit enhanced large-$z$ behavior due to nontrivial cancellations among diagrams. In Yang--Mills theory, for instance, this leads to amplitudes scaling as $z^{-1}$, which is sufficient to enable the BCFW recursion \cite{Britto:2004ap,Britto:2005fq}. 

Scalar theories display a more mysterious and intricate version of this phenomenon. Under simple shifts
\eq
p_i\rightarrow p_i+z q,\quad  p_j\rightarrow p_j-z q, 
\eqe
subject to $q\cdot p_i=q\cdot p_j=q\cdot q=0$, additional cancellations occur among \emph{all} the subleading terms: terms that individually scale as $z^k$ combine into $z^{k-1}$. Remarkably, these cancellations were shown to occur if and only if the amplitude satisfies the hidden-zero conditions associated to the pair $i,j$ \cite{Rodina:2024yfc}. For example, looking at the 5point example in Eq.~\eqref{5p_Dsubset}, we find that any term in $D_1$ scales as $z^{-2}$, while terms in $D_2$ scale as $z^{-1}$. But their sums are enhanced $D_1\sim z^{-3}$, $D_2\sim z^{-2}$.

We find this correspondence extends directly to cosmological wavefunctions. If the choice of $i,j$ splits the generating graph into exactly two subgraphs, then the $[i,j]$ shift exhibits a one-power enhancement, and this is exactly equivalent to the zero associated with that splitting. 

More generally, if $i,j$ split the graph into $V$ subgraphs---necessarily with ${\rm val}(i)=V$ and $j=n+1$---then the scaling is enhanced by $V-1$ powers $z^k \;\longrightarrow\; z^{k-(V-1)}$. This even stronger enhancement is equivalent to imposing the $V-1$ independent blob zeros associated with the multi-splitting; {\em i.e.}, any $V-1$ of the possible choices $A=V_i$, $B=\cup_{j\neq i} V_j$; the final one then follows automatically.

For example, in FIG.~\ref{fig:greenblob}, we find the $[5,8]$ shift has a one-power enhancement if and only if the zero $A=\{1,2,3,4,6,9\}$, $B=\{7\}$ is satisfied; meanwhile, the $[3,9]$ shift has a two-power enhancement if and only if any two of the following zeros are satisfied $A=\{1\}$ with $B=\{2,4,5,6,7,8\}$; $A=\{2\}$ with $B=\{1,4,5,6,7,8\}$; or $A=\{1,2\}$ with $B=\{4,5,6,7,8\}$.

It is particularly striking that BCFW recursion was originally introduced as a unitarity-based method for constructing amplitudes, whereas here the BCFW scaling itself is equivalent to a distinct recursion principle based on shuffle products, which does not assume unitarity.

\section{Outlook}
Our results point in two complementary directions. On the one hand, cosmological wavefunctions provide a natural arena in which amplitude-bootstrap ideas extend and sharpen. On the other hand, their richer combinatorial structure feeds back into flat space, revealing new structures for ordinary amplitudes: a novel interpretation in terms of factorization into shuffle products.

Our results open several directions for future research. Beyond the blob zeros studied here, rewriting the kinematic conditions $p_a\!\cdot p_b=0$ back in tube variables reveals an even broader family of non-blob cosmological zeros, greatly extending both  cosmological and ordinary amplitude zeros. Their systematic study will be explored elsewhere.

The same logic also appears to extend beyond tree level. In all examples we could check, including graphs up to three loops, zeros together with enhanced BCFW scaling uniquely determine the wavefunctions. At loop level, the tree-level shuffle product is replaced by a richer combinatorial structure, suggesting that dual factorization persists in a more general form. A systematic analysis of this loop-level structure is currently under investigation.

In this work, we also identified a relation between a known CHY formula and the kinematic/cosmological wavefunction. This opens many possible new directions. One is to identify a loop level extension. Another is to understand whether a similar CHY-type construction can generate cosmological observables for particles with spin.

Since hidden zeros, special BCFW behavior, and various uniqueness theorems arise beyond scalar theories, it is natural to ask whether dual factorization also extends to  theories like Yang-Mills or gravity.

\vskip 1cm

\noindent \emph{Acknowledgements}: 
YL would like to thank Anupam Mazumdar, Diederik Roest, and Tonnis ter Veldhuis for insightful discussions, and Eric Bergshoeff, Song He, and Karol Kampf for their support during his graduation period. We thank Shounak De for discussions, comments and earlier involvement. LR is supported by the National Natural Science Foundation of China General Program
No. 12475070 and the Beijing Natural Science Foundation International Scientist Project No. IS24014.

\section{Appendix 1: The cosmological wavefunction}\label{appendix_1}

The central observable in this work is the late-time \emph{wavefunction of the universe}. Conceptually, it is the cosmological analogue of scattering data: rather than relating in- and out-states, it assigns an amplitude to a field configuration on a final time slice. Once the wavefunction is known, equal-time correlators follow from Born's rule by integrating against \(|\Psi|^2\). In this sense, the wavefunction coefficients play for cosmology a role closely analogous to that played by scattering amplitudes in flat space.

We consider a conformally-coupled scalar with polynomial interactions in an FRW universe,
\begin{align}
    \mathcal{S} = \int \td^d x \, \td\eta \sqrt{-g} \left[-\frac{1}{2} g^{\mu \nu} \partial_{\mu} \phi \partial_{\nu} \phi - \frac{1}{2} \xi R \phi^2 - \sum_{k \geq 3} \frac{\lambda_k}{k!} \phi^k \right],
    \label{eq:actioninFRW_rewrite_full}
\end{align}
with FRW metric \( \td s^2 = a^2(\eta)\left[-\td\eta^2+\td\vec{x}^{\,2}\right]\). For the conformal value \(\xi=\frac{d-1}{4d}\), this theory is equivalent to a flat-space scalar with time-dependent couplings,
\begin{align}
    S[\phi] = \int \td^d x \, \td\eta \left[-\frac{1}{2}(\partial\phi)^2 - \sum_{k\ge 3}\frac{\lambda_k(\eta)}{k!}\phi^k\right].
    \label{eq:actioninflatspace_rewrite_full}
\end{align}
In this paper we focus on the constant-coupling case \(\lambda_k(\eta)=\text{const.}\), since more general cosmological observables may be recovered from it by a simple integral transform \cite{Arkani-Hamed:2023kig,Arkani-Hamed:2023bsv,De:2023xue,De:2024zic}.

The late-time vacuum wavefunction is defined by the path integral
\begin{align}
    \Psi[\Phi]
    =
    \int_{\phi(-\infty(1-i\epsilon))}^{\phi(0)=\Phi}\mathcal D\phi\,e^{iS[\phi]} \, ,
    \label{eq:Uniwavefunc_rewrite_full}
\end{align}
where the field approaches the adiabatic/Bunch--Davies vacuum in the far past and is fixed to the boundary value \(\Phi(\vec x)\) at late times. In momentum space one may expand
\begin{align}
    \Psi[\Phi]
    =
    \exp\!\bigg[
    &\sum_{n\ge2}\frac{1}{n!}
    \int \prod_{i=1}^n \td^d \vec{k}_i\,
    \psi_n(\vec{k}_1,\ldots,\vec{k}_n)\,
    \nonumber\\
    &\Phi(\vec{k}_1)\cdots \Phi(\vec{k}_n)\,
    \delta^{(d)}\!\left(\sum_{i=1}^n \vec{k}_i\right)
    \bigg] ,
\end{align}
and the coefficients \(\psi_n\) determine equal-time correlators through
\begin{align}
    \langle \Phi(\vec{k}_1)\cdots \Phi(\vec{k}_n)\rangle
    \propto
    \int \mathcal D\Phi\,
    \Phi(\vec{k}_1)\cdots \Phi(\vec{k}_n)\,
    |\Psi[\Phi]|^2 .
\end{align}
For the questions studied here, it is more convenient to organize perturbation theory by truncated interaction graphs \(\G\) rather than by the number of external boundary insertions. We therefore write \(\psi_{\G}\) for the contribution associated with a graph \(\G\).

At tree level, \(\psi_{\G}\) may be computed directly from the usual time-integral representation, with one bulk time \(\eta_v\) for each vertex \(v\) of \(\G\). Schematically,
\begin{align}
    \psi_{\G}(\{x_v\},\{y_e\})
    =
    \int_{-\infty}^0 & \prod_{v\in V(\G)} \td\eta_v\,
    e^{i x_v \eta_v}\nonumber\\
    & \prod_{e=(v\leftrightarrow v')\in E(\G)}
    G_{y_e}(\eta_v,\eta_{v'}) \, ,
    \label{eq:time_integral_schematic}
\end{align}
where \(x_v\) is the sum of the external energies entering vertex \(v\), and \(y_e=|\vec p_e|\) is the energy carried by the internal edge \(e\). These variables are naturally \emph{non-Lorentz-invariant}: unlike scattering amplitudes, the wavefunction is defined relative to a preferred late-time slice, so only spatial momentum is conserved. It is precisely these energy variables \(x_v\) and \(y_e\) that enter the combinatorial description below.

A remarkable feature of the wavefunction coefficients is that, for tree graphs, the time integrals can be replaced by a purely combinatorial formula \cite{Arkani-Hamed:2017fdk}. To each connected subgraph, or \emph{tube}, \(\tau\subset \G\), one associates the energy
\begin{align}
    S_\tau
    :=
    \sum_{v\in \tau} x_v
    +
    \sum_{e\in \partial\tau} y_e \, ,
\end{align}
where \(\partial\tau\) denotes the set of edges leaving \(\tau\). The coefficient \(\psi_{\G}\) is then given by a sum over maximal tubings \(T\) of the graph,
\begin{align}
    \psi_{\G}
    =
    \sum_{T}\prod_{\tau\in T}\frac{1}{S_{\tau}}
    =
    \frac{1}{\Stot\mathop{\prod}\limits_{\tau\in \{\text{1-tubes}\}}S_{\tau}}\,
    \widetilde{\psi}_{\G} \, ,
    \label{eq: wavefunction coefficients_rewrite_full}
\end{align}
where \(\Stot=\sum_{v}x_v\) is the total energy, and \(\widetilde{\psi}_{\G}\) denotes the stripped coefficient after removing the universal total-energy and 1-tube factors. This stripped object will be our main focus throughout the paper.

As an example, for the \(222\)-star one finds
\begin{align}
    \label{222_rewrite_full}
\widetilde{\psi}_{\G}
=
&\frac{1}{S_{34}S_{134}}
+\frac{1}{S_{13}S_{134}}
+\frac{1}{S_{23}S_{132}}\nonumber\\
&
+\frac{1}{S_{13}S_{132}}
+\frac{1}{S_{23}S_{234}}
+\frac{1}{S_{34}S_{234}} \, .
\end{align}
This six-term expression is simply the sum over the six maximal tubings of the star graph.

Finally, recent work showed that the stripped coefficients \(\widetilde{\psi}_{\G}\) vanish on several classes of loci, including parametric, wavefunction, and factorization zeros~\cite{De:2025bmf}. Below  we will show more explicitly how the blob zeros extend these previously discovered zeros, and unify under the same simple structure.

\subsection{Two maps from cosmology to amplitudes}
It is important to distinguish two different maps from wavefunctions to flat-space quantities. The standard flat-space limit is obtained by taking the residue at vanishing total energy,
\[
\Stot=\sum_v x_v \to 0 .
\]
For a single graph \(\G\), this residue extracts exactly the flat-space Feynman diagram of the original \(\phi^n\) theory associated with \(\G\). In particular, graphs with higher-valency vertices map to the corresponding \(\phi^n\) diagrams, not to sums of color-ordered cubic amplitudes. This residue map is therefore entirely distinct from the kinematic map introduced in the main text. The latter acts on the stripped wavefunction coefficient \(\widetilde{\psi}_{\G}\) by replacing tube energies with Mandelstam-like invariants, and produces generalized ordered \(\Tr(\phi^3)\)-type objects instead.

\section{Appendix 2: Unification of cosmological zeros}
\label{appendix_2}
The blob zeros introduced in this work unify all previously known cosmological zeros \cite{De:2025bmf}.
Concretely, we find:
\begin{itemize}
\item Parametric zeros (valency $2$), $S_a = S_{\text{tot}} = 0$ $\;\Leftrightarrow\;$ blob zero with $i=a$, $j=n{+}1$.
\item Wavefunction zeros of a chain graph $(12,\ldots,n)$ $\;\Leftrightarrow\;$ blob zero with $i=1$, $j=n$.
\item Factorization zeros of chain subgraphs bounded by $a_L$ and $a_R$ $\;\Leftrightarrow\;$ blob zero with $i=a_L$, $j=a_R$.
\item Parametric zeros for vertices of valency $|v|>2$ arise as composite conditions. Choosing $i=v$ and $j=n{+}1$ partitions the graph into $|v|$ branches $V_1,\dots,V_{|v|}$, and the parametric zero is equivalent to imposing the pairwise constraints
\[
p_a \cdot p_b = 0 \,, \qquad a \in V_r,\; b \in V_s,\; \textrm{for all }r \neq s \,,
\]
\end{itemize}

This makes clear that all previously known cosmological zeros are unified as blob zeros. 

In addition to these known zeros, the new zero type we find are the blob zeros that can be imposed on a \emph{single branch},
\[
p_a \cdot p_b = 0 \,, \qquad a \in V_r,\; b \in \bigcup_{s\neq r} V_s \,,
\]
which are stronger than constraints for the cosmological parametric zero. These new zeros are mandatory to  prove uniqueness for arbitrary graphs, and therefore complete the structure.

\section{Appendix 3: Proof that zeros uniquely fix the $D$-subsets}
\label{appendix_3}

In this Appendix we first review the amplitude proof that a zero condition fixes the coefficients within each corresponding $D$-subset. We then extend the proof for the more general wavefunctions. Combined with the construction in the main text, this establishes the equivalence between the zero condition and the associated shuffle product.

Any local ansatz, such as Eq.~\eqref{5pt}, may be decomposed into $D$-subsets,
\begin{align}
\mathcal{B}_n \;=\; \sum_\alpha D_\alpha \, ,
\end{align}
where the decomposition depends on the chosen zero. Each $D_\alpha$ is a set of skeleton diagrams related by the elementary moves compatible with that zero condition.  For example, for the five-point zero $\{1\},\{3,4\}$, one finds
\eqa
D_1 &=& \frac{c_1}{s_{12}s_{123}}+\frac{c_2}{s_{23}s_{123}}+\frac{c_4}{s_{23}s_{234}} \, ,\\
D_2 &=& \frac{c_3}{s_{12}s_{34}}+\frac{c_5}{s_{34}s_{234}} \, .
\eqae
with the diagrams shown in FIG.\ref{fig: 5pt D-subsets}.
Different zeros lead to different decompositions, and initially each term in each $D_\alpha$ carries an independent coefficient.

The key result of Ref.~\cite{Rodina:2024yfc} is that the zero holds if and only if two conditions are satisfied:
\begin{itemize}
\item each $D_\alpha$ vanishes independently on the zero locus;
\item all coefficients within a given $D_\alpha$ are equal.
\end{itemize}
In the example above, this means that $\mathcal{B}_5$ satisfies the zero precisely when
\[
c_1=c_2=c_4,
\qquad
c_3=c_5.
\]

The proof proceeds by taking successive residues compatible with the zero condition. First, residues isolating a given $D$-subset show that each $D_\alpha$ vanishes individually. In the above case, this goes as follows. First take the residue on $s_{34}\rightarrow 0$. This uniquely picks $D_2$, and implies that $\textrm{Res}[D_1|_{\rm zero}]=0$. Then one shows this condition also holds away from the residue, so that $D_1|_{\rm zero}=0$. Since $(D_1+D_2)|_{\rm zero}=0$ by assumption this automaticaly implies $D_2|_{\rm zero}=0$. When more D-subsets are present, the argument follows several such steps.

One then takes further residues on boundary propagators to compare neighboring terms inside a given $D$-subset, namely terms differing by a single elementary move. The zero condition then forces their coefficients to agree. In the example above, once $D_1|_{\rm zero}=0$, taking the residue $s'_{123}\to 0$ gives
\begin{align}
0
=
\res{s'_{123}\to 0}
\left[
\left.
\left(
\frac{c_1}{s_{12}s_{123}}+\frac{c_2}{s_{23}s_{123}}
\right)
\right|_{\rm zero}
\right]
=
\frac{c_1}{s_{12}}+\frac{c_2}{-s_{12}} \, ,
\end{align}
where on the zero condition $s_{123}\rightarrow s'_{123}=s_{12}+s_{23}$, and on the combined zero-plus-residue locus one has $s_{23}=-s_{12}$. This fixes $c_1=c_2$. A similar residue then gives $c_2=c_4$, and hence all coefficients in $D_1$ are equal. The same reasoning applied to $D_2$ yields $c_3=c_5$.

The argument is straightforward in principle, but one important point must be respected throughout: the residues used in the proof must be compatible with the zero condition. Otherwise one may inadvertently introduce degeneracies and pick up unwanted terms. Equivalently, every residue must commute with the imposed zero. This compatibility condition is the essential ingredient in extending the proof from ordinary amplitudes to the cosmological case.

\subsection{Cosmological extension}
In the cosmological case, the only new ingredient is the presence of additional orderings. These make the individually-vanishing subset bulkier, since the kinematic wavefunction includes amplitudes of different orderings, and there exist $D$-subsets sharing the same non-boundary propagators but differing in ordering. We denote this bulkier subset as a $\hat{D}$-subset. From this set-up, the proof of uniqueness proceeds in two steps.
\begin{itemize}
\item A zero requires individually vanishing $D$-subsets 
\item Each $D$-subset is uniquely fixed
\end{itemize}
The first step uses the standard BCFW scaling plus ``away-from-residue'' argument. The residue of $\mathcal{A}_G$ on all non-boundary propagators of a $\hat{D}$-subset isolates this subset, and vanishes due to the zero condition. Since the non-boundary propagators scale differently under BCFW shifts, they cannot be expressed as linear combinations of only boundary propagators. Hence, the vanishing of a $\hat{D}$-subset holds for general kinematics away from these residues; in other words, each $\hat{D}$-subset vanishes individually on the zero loci.

In the second step, residues of $\hat{D}$ are taken on boundary propagators to isolate two ``neighboring'' terms, which share all but one propagator. Iterating this pairwise procedure relates coefficients pairwise. Interestingly, such residues isolate only terms with the same amplitude ordering in $\hat{D}$, since Feynman diagrams from different orderings differ by at least two boundary propagators. Therefore, by taking residue of $\hat{D}$ on boundary propagators, we establish the uniqueness of $D$-subset in each $\hat{D}$, with free coefficients remained among different orderings of $\hat{D}$.

A potential subtlety arises in this step. While residues on non-boundary propagators are unaffected by the zero condition, taking residues on boundary propagators requires that the operations of taking the zero and taking residues commute, as mentioned above; otherwise, the isolation of two terms would fail. In the kinematic wavefunction, the sum over amplitude orderings could in principle obstruct this commutativity. In the following, we show that this does not occur.

In the skeleton diagrams associated with a given zero, boundary propagators take the form $s_{A',j}$, $s_{B',j}$, or $s_{A',B',j}$. Commutativity can fail if the zero collapses propagators so that distinct propagators become identical after taking residues. Such collapses arise from kinematic identities among inverse propagators. The only potentially dangerous identity is
\[
s_{A',B',j} = s_{A',j} + s_{B',j}\,,
\]
evaluated on the zero $p_{a\in A}\!\cdot p_{b\in B}=0$, with $A'\subset A$ and $B'\subset B$. If the residues are taken in the sequence,
\begin{align}
    \cdots \res{s_{A',B',j}\to 0}\res{s_{A',j}\to 0} \hat{D}\, ,
\end{align}
then $s_{A',B',j}\to s_{B',j}$ on the zero with $s_{A',j}\to 0$, so that evaluating $\hat{D}$ on the zero and taking $\res{s_{A',B',j}\to 0}$ do not commute, due to an extra contribution from $1/s_{B',j}$. However, this contribution is eliminated after taking $\res{s_{A',j}\to 0}$, since $1/s_{A',j}$ and $1/s_{B',j}$ are incompatible poles and cannot appear in the same diagram, regardless of ordering. Therefore, the commutativity between imposing the zero and taking $\res{s_{A',B',j}\to 0}$ is secured.

Residues on other types of boundary propagators do not introduce any risk of non-commutativity. Consider $s_{A',B',j}=s_{A_1,A_2,B_1,B_2,j}$, where $A'=A_1\cup A_2\subset A$ and $B'=B_1\cup B_2\subset B$. Suppose the residues are taken in the sequence
\begin{align}
    \cdots  \res{s_{A',B',j}\to 0}
    \res{s_{A_1,B_1,j}\to 0}\hat{D} \,\, .
\end{align}
On the zero locus, the kinematic identity yields
\[
s_{A',B',j} \to s_{A_1,B_1,j} + s_{A_2,B_2,j} + p_{A_1}\!\cdot p_{A_2} + p_{B_1}\!\cdot p_{B_2}\,,
\]
which is even less likely to induce non-commutativity, as the additional terms $p_{A_1}\!\cdot p_{A_2} + p_{B_1}\!\cdot p_{B_2}$ prevent any collapse of propagators. This completes the proof of commutativity in the presence of multiple amplitude orderings within $\hat{D}$.

\subsection{Factorization and counting examples}
In this Appendix we present several non-trivial examples for amplitudes or kinematic wavefunctions are uniquely fixed via zeros.

\begin{figure}[h!]
    \centering
    \includegraphics[width=1\linewidth]{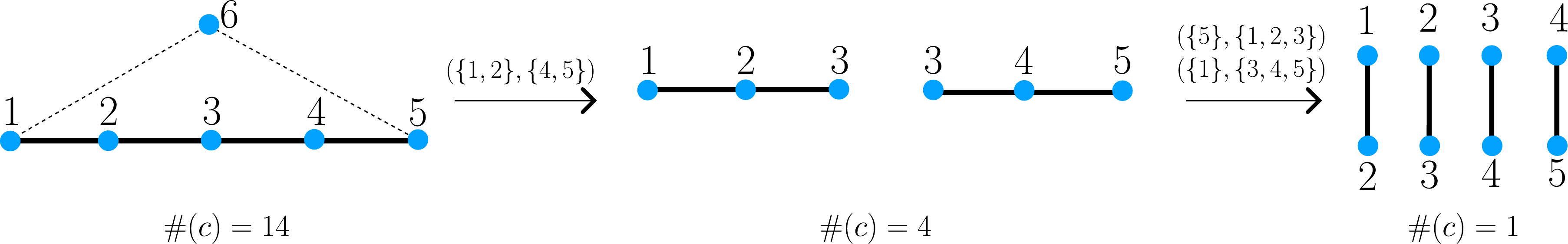}
    \caption{The reduction of 5-chain via imposing the blob zeros. At each step, the subgraphs are listed.}
    \label{fig: 5-chain graph reduction}
\end{figure}

\noindent\textbf{5-chain (6-point amplitude)} The first example is the 5-chain or equivalently the 6-point amplitude. In FIG.~\ref{fig: 5-chain graph reduction}, the zeros are imposed subsequently and eventually factorize the 5-chain into four 2-chain seeds, which has a overall normalization fixed the the coupling as mentioned in the main text.

\begin{figure}[h!]
    \centering
    \includegraphics[width=1\linewidth]{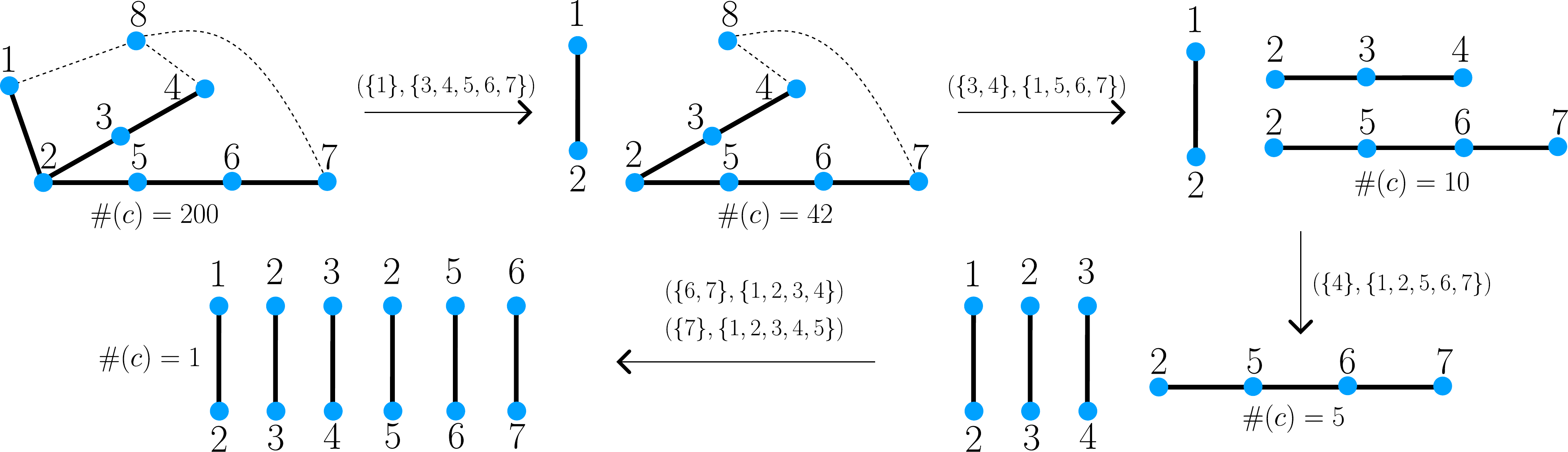}
    \caption{The reduction of 234-star via imposing the blob zeros. At each step, the subgraphs are listed.}
    \label{fig: 234 graph reduction}
\end{figure}

\noindent\textbf{234-star}  An example that goes beyond the 222-star that we have shown in the main text, is the 234-star. In FIG.~\ref{fig: 234 graph reduction}, the zeros are imposed subsequently and eventually factorize the 234-star into six 2-chain seeds, which has a overall normalization fixed the the coupling.

\begin{figure}[h!]
    \centering
    \includegraphics[width=1\linewidth]{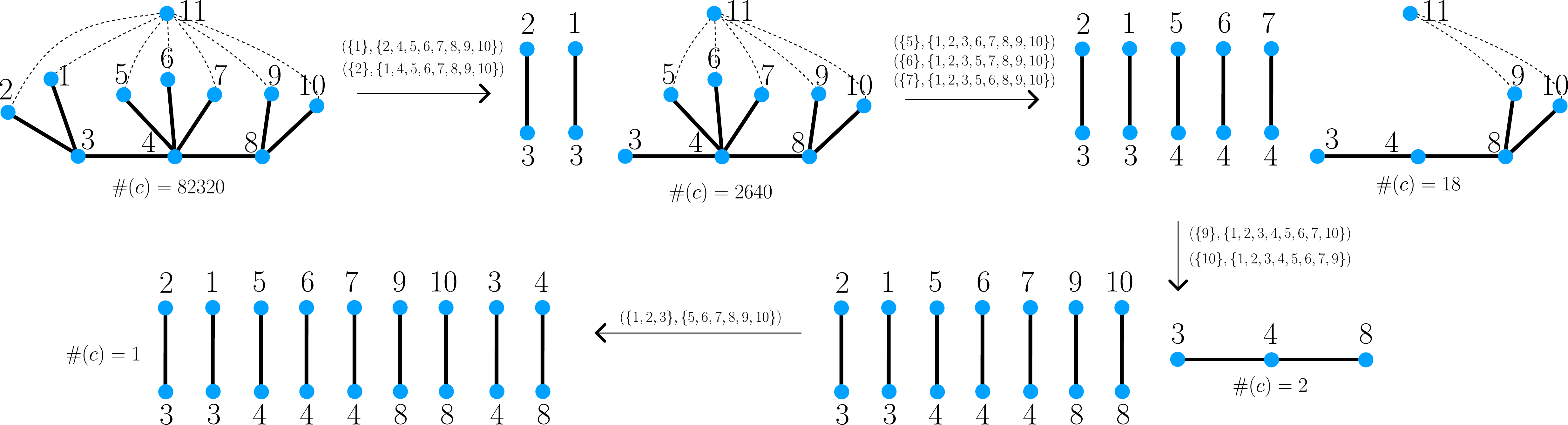}
    \caption{The reduction of 222-2222-222-star via imposing the blob zeros. At each step, the subgraphs are listed.}
    \label{fig: 222-2222-222 graph reduction}
\end{figure}

\noindent \textbf{222-2222-222 triple star} The most complicated example we show is the following star with triple centers. The reduction process via imposing blob zeros subsequently is illustrated by FIG.~\ref{fig: 222-2222-222 graph reduction}. At each steps, the number of coefficients is the product of all subgraphs, \`a la \eqref{factorN}. As shown, the graph is eventually decomposed into a nine 2-chain seeds, which has a overall normalization fixed the the coupling. Remarkably, 8 zero conditions are sufficient to fully fix the starting 82320 term ansatz!

\section{Appendix 4: Near-zero factorization from dual factorization}
\label{appendix_4}

In this Appendix we illustrate how near-zero factorization follows directly from dual factorization. In particular, once an amplitude is written as a shuffle product associated with a zero condition, taking a near-zero limit turns that shuffle decomposition into the familiar product form. As an immediate corollary, the same conclusion applies to cosmological wavefunctions.

Consider a zero condition of the form
\[
p_{a\in A}\!\cdot p_{b\in B}=0 .
\]
It was shown in Ref.~\cite{Arkani-Hamed:2023swr} that if one relaxes a single condition, say $c_{mn}\equiv -p_m\!\cdot p_n\neq 0$, then the amplitude develops the near-zero factorized form
\begin{equation}
A\Big|_{\text{near-zero}}
\;\longrightarrow\;
\frac{c_{mn}}{X_T X_B}\times
\widetilde{A}_T\times
\widetilde{A}_B \, ,
\end{equation}
where the lower-point amplitudes are evaluated on deformed kinematics determined by the chosen $c_{mn}$.

Our point is that this structure is already encoded in the shuffle decomposition itself. More precisely, each individual $D$-subset --- the basic building block associated with a given zero --- not only satisfies the exact zero condition, but also factorizes on the corresponding near-zero locus.

Let us begin with the five-point example of Eq.~\eqref{5p_Dsubset}, with zero $A=\{1\}$, $B=\{3,4\}$. We take the near-zero limit in which $p_1\!\cdot p_3=0$ while $p_1\!\cdot p_4\neq 0$. Then the two $D$-subsets factorize as
\begin{align}
D_1 &\;\longrightarrow\; \frac{c_{14}}{s'_{12}s'_{234}}\frac{1}{s'_{23}} \, ,\\
D_2 &\;\longrightarrow\; \frac{c_{14}}{s'_{12}s'_{234}}\frac{1}{s'_{34}} \, ,
\end{align}
where $s'\equiv s|_{\text{near-zero}}$, some momenta being off-shell and some propagators being massive \`a la \cite{Cao:2024gln,Cao:2024qpp} and \cite{Naculich:2015zha}. Summing them gives
\begin{align}
A_5=D_1+D_2
&\;\longrightarrow\;
\frac{c_{14}}{s'_{12}s'_{234}}
\left(
\frac{1}{s'_{23}}+\frac{1}{s'_{34}}
\right)
\nonumber\\
&=
(\text{prefactor})\times
\widetilde{A}_3(512)\times
\widetilde{A}_4(2345)\, ,
\end{align}
where we have introduced the trivial three-point factor $\widetilde{A}_3$. Thus we see the near-zero factorization follows directly from the structure of the $D$-subsets.

The same pattern persists at higher point. At six points, for the zero
$A=\{1,2\}$, $B=\{4,5\}$, the amplitude decomposes into four $D$-subsets, each arising from the shuffle of the two lower-point factors. Taking a near-zero limit, for example with $c_{25}\neq 0$, one finds that each $D$-subset again factorizes:
\begin{align}
D_{1,2,3,4}\;\longrightarrow\;
\left\{
F\frac{1}{s'_{12}}\frac{1}{s'_{34}},\ 
F\frac{1}{s'_{16}}\frac{1}{s'_{34}},\ 
F\frac{1}{s'_{12}}\frac{1}{s'_{45}},\ 
F\frac{1}{s'_{16}}\frac{1}{s'_{45}}
\right\},
\end{align}
where $F=\frac{c_{25}}{s'_{123}s'_{345}}$. Schematically, each subset takes the form
\begin{equation}
D \;\longrightarrow\; (\text{prefactor})\times
\widetilde{\Gamma}_A\times
\widetilde{\Gamma}_B \, ,
\end{equation}
and summing all four pieces gives
\begin{equation}
A_6
\;\longrightarrow\;
(\text{prefactor})\times
\widetilde{A}_4(6123)\times
\widetilde{A}_4(3456)\, .
\end{equation}

The general pattern is therefore clear. Given a zero on $(A,B)$, dual factorization implies the exact shuffle form
\begin{equation}
A(i,A,j,B)
=
A(i,A,j)\dsqcup A(i,B,j)\, .
\end{equation}
Upon taking a near-zero limit with one cross-invariant $c_{mn}\neq0$, this shuffle decomposition collapses into an ordinary product,
\begin{equation}
A(i,A,j,B)
\;\longrightarrow\;
\frac{c_{mn}}{s'_{A,j}\,s'_{j,B}}\,
\widetilde{A}(i,A,j)\,
\widetilde{A}(i,B,j)\, ,
\end{equation}
with the lower-point factors evaluated on the corresponding near-zero kinematics.

In summary, near-zero factorization follows directly from dual factorization. The exact zero is encoded by a shuffle product, while relaxing one zero condition turns that shuffle product into an ordinary product of lower-point objects. Since the same shuffle structure governs cosmological wavefunctions, the same conclusion applies there as well.

\bibliography{0latest}

\end{document}